\documentclass[twocolumn]{revtex4-1}

\newcommand{\be}{\begin{equation}}
\newcommand{\ee}{\end{equation}}

\usepackage{amsmath}
\usepackage{amssymb}
\usepackage{amsthm}
\usepackage{pstricks}
\usepackage{graphicx}

\begin{document}

\title{Doping of Si nanoparticles: the effect of oxidation}%

\author{A.~Carvalho}
\email{aicarvalho@ua.pt}
\affiliation{Department of Physics, I3N, University of Aveiro, Campus Universit\'ario de Santiago, 3810-193 Aveiro, Portugal}
\author{S. \"Oberg}
\affiliation{Department of Engineering Sciences and Mathematics, Lule{\aa} University of Technology, Lule{\aa}~S-97187, Sweden}
\author{M. Barroso}
\affiliation{Department of Physics, I3N, University of Aveiro, Campus Universit\'ario de Santiago, 3810-193 Aveiro, Portugal}
\author{M. J. Rayson}
\affiliation{Department of Engineering Sciences and Mathematics, Lule{\aa} University of Technology, Lule{\aa}~S-97187, Sweden}
\author{P. R. Briddon}
\affiliation{Electrical, Electronic and Computer Engineering, University of Newcastle upon Tyne, Newcastle upon Tyne NE1 7RU, United Kingdom}

\begin{abstract}
The preferred location of boron and phosphorus in oxidized free-standing Si nanoparticles was
investigated using a first-principles density functional approach. 
The calculated formation energies indicate that P should segregate to the silicon core,
whereas B is equally stable in the Si and SiO$_2$ regions.
Our models thus suggest that, in contrast with nanocrystals with H-terminated surfaces,
the efficiency of phosphorus incorporation in oxidized Si nanoparticles can
be improved by thermal annealing.
\end{abstract}


\date{\today}%

\maketitle   

Silicon nanocrystals have emerged as a promising strategy to achieve light generation and amplification
using silicon.
Adding to the advantages of bulk silicon, they offer variable bandgap, increased luminescence efficiency\cite{takeoka-PRB-62-16820},
and the possibility of multiple exciton generation\cite{timmerman-NN-6-710,kovalev-PRL-81-2803}.
Together with size, shape, and surface functionalization, impurity doping can be used
to control the optical and electronic properties,
conferring to the silicon nanocrystals an extraordinary degree of design flexibility.

Boron and phosphorus doping has been achieved both for self-standing nanoparticles 
synthesized using plasma approaches\cite{pi-APL-92-123102,stegner-PRB-80-165326},
and for Si nanoparticles embedded in glass matrix (usually SiO$_2$, borosilicate or phosphosilicate)\cite{fuji-in-pavesi-book}.
However, the doping efficiency is often limited by the segregation of the impurity to the surface
or to the surrounding matrix\cite{pi-APL-92-123102,cantele-PRB-72-113303,carvalho-OH,stegner-PRB-80-165326}.

The tendency for impurity segregation depends on the nanoparticle morphology and growth method.
In Si nanocrystals with hydrogen-terminated surface, 
according to both experimental and theoretical studies,
P segregates to the surface, saturating Si dangling bonds\cite{pi-APL-92-123102,ma-APL-98-173103,carvalho-OH}.
Boron was found by theoretical studies\cite{xu-PRB-75-235304,cantele-PRB-72-113303,carvalho-OH} to be more stable in the sub-surface layer
or at the surface itself,
but in practice seems to be incorporated in the nanocrystal core, since subsequent oxidation of the 
outer shells of the nanocrystal does not decrease the fraction of electrically active B atoms\cite{pi-APL-92-123102}.

The opposite behavior is observed in SiO$_2$-embedded silicon nanocrystals. 
In this case, phosphorus segregates to the Si-rich region\cite{perego-Nt-21-02560}. 
Segregation to the silicon region has also been observed in Si-SiO$_2$ interfaces,\cite{grove-JAP-35-2695}
where it has been confirmed by atomistic modeling\cite{cole-SS-601-4888}.
For heavily implanted material, though, P pile-up at the interface is observed instead\cite{chang-JAP-103-053517}.
The case of boron is not so clear. 
It has been found to have a segregation coefficient $s=C_{\rm Si}/C_{\rm SiO_2}=0.3$, 
where $=C_{\rm Si}$ and $C_{\rm SiO_2}$ are the equilibrium concentrations\cite{grove-JAP-35-2695}.
The piling-up at the Si-side of the interface during segregation was justified by
its low diffusivity in SiO$_2$.
However, theoretical results indicate that it is more stable at the Si-side of the interface or in the Si bulk.\cite{furuhashi-ELEX-1-126,jun-ME-89-120,cole-SS-601-4888}

In this paper, we use first principles calculations, based on density functional theory,
to show that in Si nanocrystals covered by a SiO$_2$ shell,
the preferred location of P is at the nanocrystal core, whereas B has similar 
energy on either side, thus confirming that the behavior is different from H-passivated
silicon nanocrystals.

The calculations were carried out using the {\sc Aimpro} code\cite{briddon-pssb-217-131,rayson-cpc-178-128,rayson-PRB-80-205104,briddon-pssb-248-1309}.
The core electrons were modeled with the dual space separable pseudopotentials by Hartwigsen, Goedecker and Hutter~\cite{HGH}.
The basis set consisted of atom-centered Cartesian-Gaussian functions~\cite{goss-TAP-104-69}.
A contracted basis with 13 functions per atom was used for Si, 
whereas for O and B we used uncontracted basis sets with a total of 40 and 28 Gaussian basis functions per atom, respectively. 
A Pad\'e parametrization of a functional based on the
local density approximation was employed for the
exchange and correlation energy\cite{pade}.

\begin{figure}[htb]%
\includegraphics{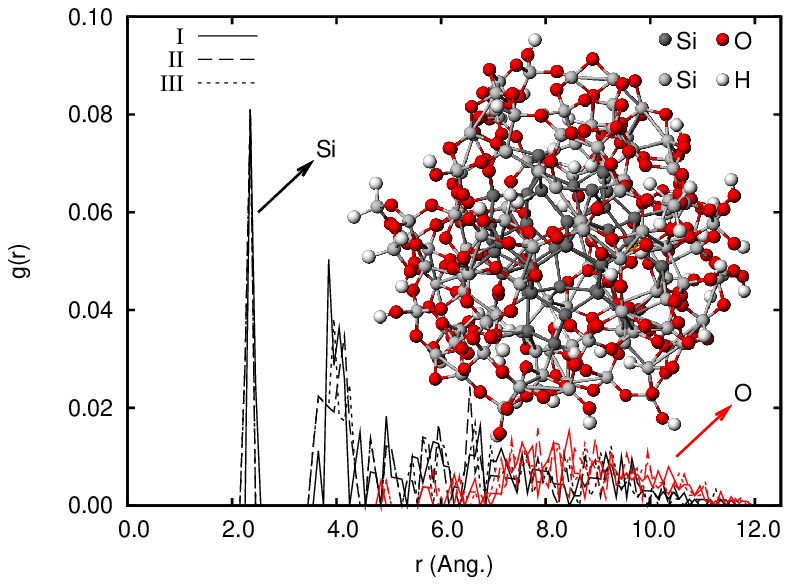}
\caption{
Radial pair distribution functions for nanoparticles I-III, 
calculated relative to the central Si atom, for the Si species (black lines)
and O species (lighter line).
Inset: Structure of an undoped nanoparticle (nanoparticle I).
Silicon atoms in the core region are represented in a darker shade of gray.
}
\label{fig:struc}
\end{figure}

The models for the oxidized silicon nanoparticles consisted of
an approximately spherical Si core of 1.5~nm diameter, 
surrounded by an amorphous SiO$_2$ shell with about 2~nm outer diameter.
Three models were considered.
The starting structures were obtained from the results of molecular dynamics simulations
of 1800~K annealings during 2.0 ps, 2.3 ps and 2.6 ps,
as detailed in Ref.~\onlinecite{nanoSi-EMRS}.
Dangling bonds were passivated with hydrogen atoms.
The compositions of the Si-NPs were Si$_{161}$O$_{196}$H$_{59}$ for nanoparticles I and II, and
Si$_{161}$O$_{196}$H$_{57}$ for nanoparticle III.
Nanoparticle I was the same used in a previous study\cite{nanoSi-EMRS}.
The models had no periodic boundary conditions imposed. 
The starting structures were relaxed at 0 K, using a conjugate gradient algorithm,
with the {\sc Aimpro} code.
The radial pair distribution function for Si, shown in Fig.~\ref{fig:struc} for the
three nanocrystal models, shows that the core consisting of the first two shells of atoms 
surrounding the central Si atom has crystalline character, whereas the SiO$_2$ is amorphous.

The preferred locations for substitutional boron (B$_s$) and substitutional phosphorus (P$_s$)
were investigated by comparing the energy of the doped nanoparticles obtained by
replacing boron or phosphorus for each of the 161 silicon atoms in the three models I-III.
We have considered the equilibrium charge states of B$_s$ and P$_s$, 
which are B$_s^-$ and P$_s^+$ both in Si and SiO$_2$\cite{previous}.
The formation energies ($E_f$) of B$_s^-$ and P$_s^+$ are given 
relative to those of the equivalent defects in bulk silicon,
calculated using as standard 512 atom supercell:
$$E_f^i=E[{\rm NP}^i:X]-E[{\rm NP}]-\{ E[ {\rm Si_{511}}:X ]-E[{\rm Si}_{512}] \},$$
where `NP' is the undoped silicon nanoparticle and 
\mbox{NP$^i:X$} is a similar nanoparticle but with $X={\rm B}^-$ or P$^+$ at the $i$th Si site.
The energy scales of both systems have been aligned using the vacuum energy as standard.

\begin{figure}[htb]%
\includegraphics[height=8.6cm,angle=-90]{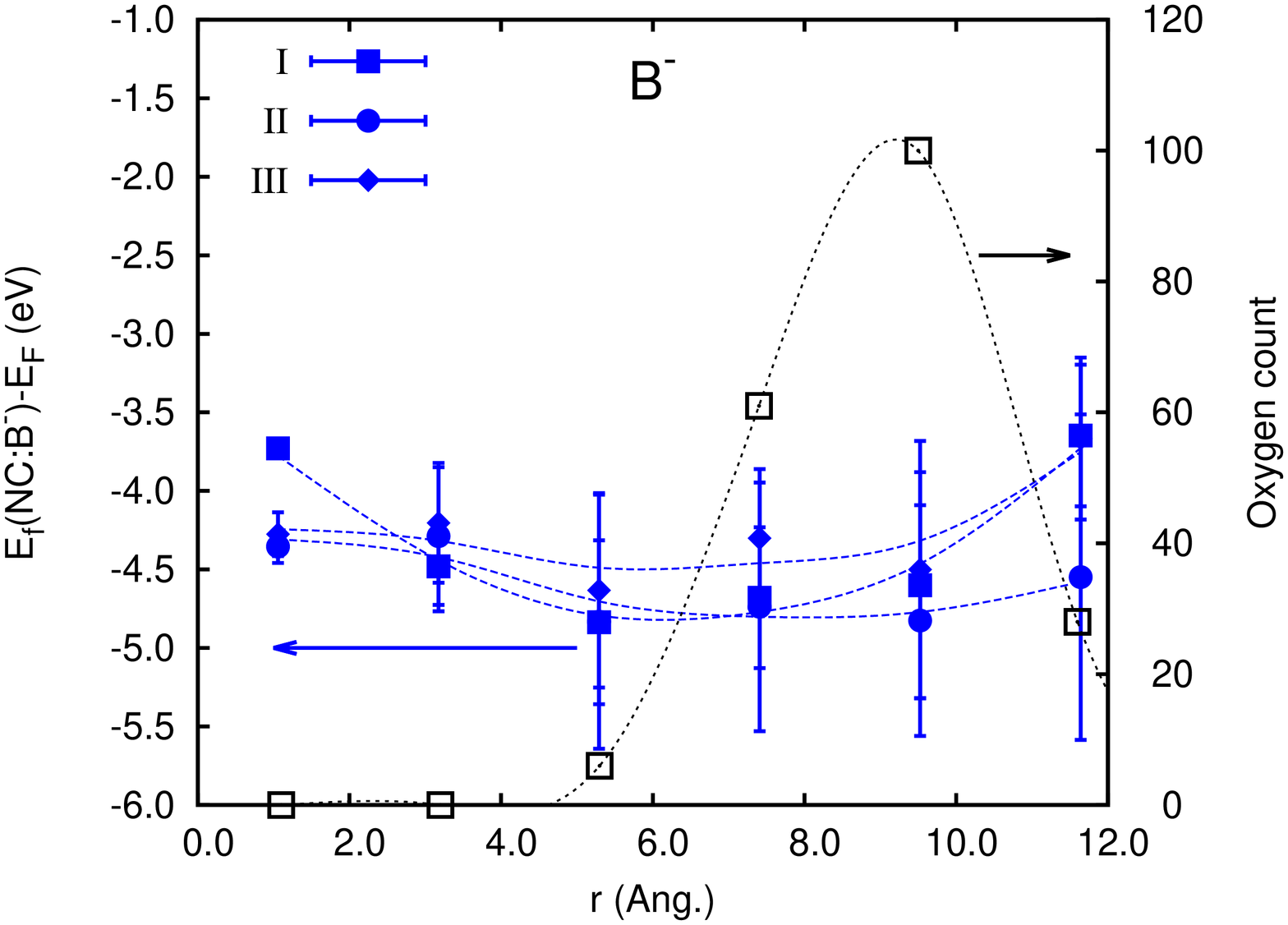}
\includegraphics[height=8.6cm,angle=-90]{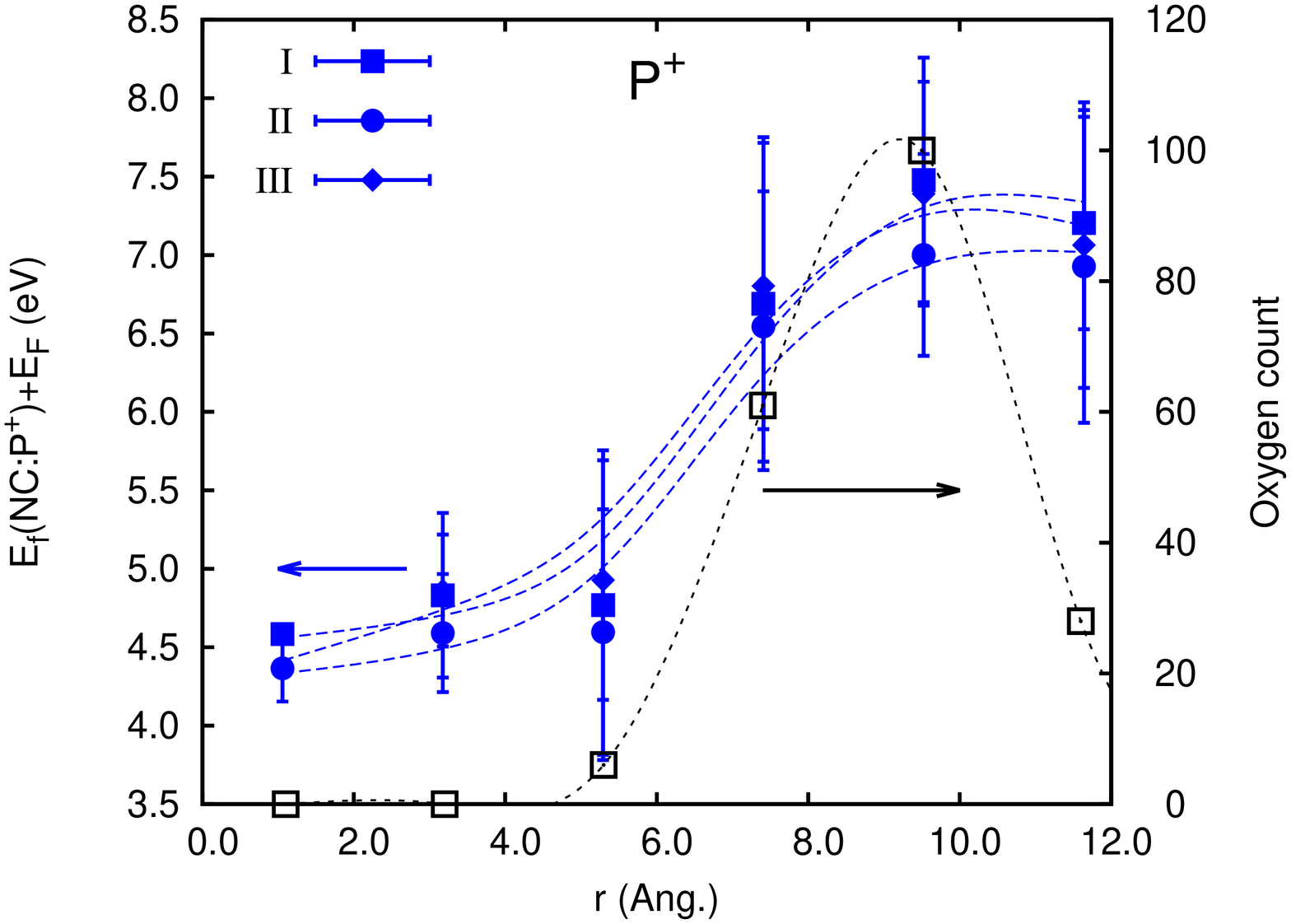}
\caption{Formation energy of the SiO$_2$-covered nanoparticle
doped with B (top) or P (bottom),
as a function of the distance from the center.
The filled symbols correspond to average formation energies for distances
between $r-2$~\AA\ and $r+2$~\AA, and
their errorbars represent the statistic standard deviation of the samples.
Open squares represent the distribution of oxygen (coincident for I, II and III).
}
\label{fig:plot}
\end{figure}

The results show an appreciable difference between B$_s$ and P$_s$ (Fig.~\ref{fig:plot}).
For boron, the formation energy is nearly independent on the location.
The average formation energies are slightly lower at the interface,
but the difference between average values is smaller than the width of the energy distributions represented by the errorbars.
Phosphorus, however has a preference for the Si core region.
At the oxide shell, the formation energy is higher by about 2~eV.

\begin{figure}[htb]%
\includegraphics[height=8.6cm,angle=-90]{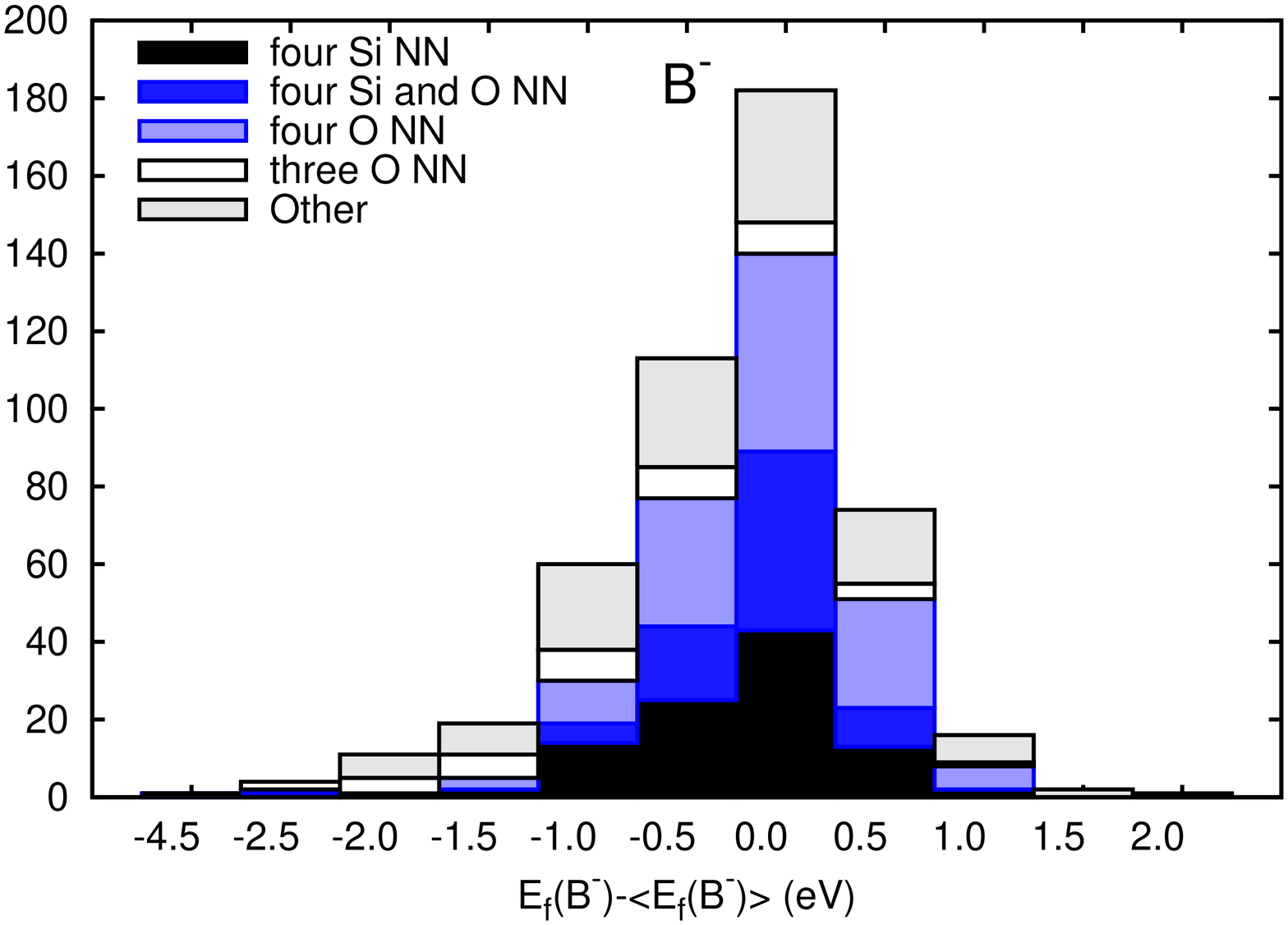}
\includegraphics[height=8.6cm,angle=-90]{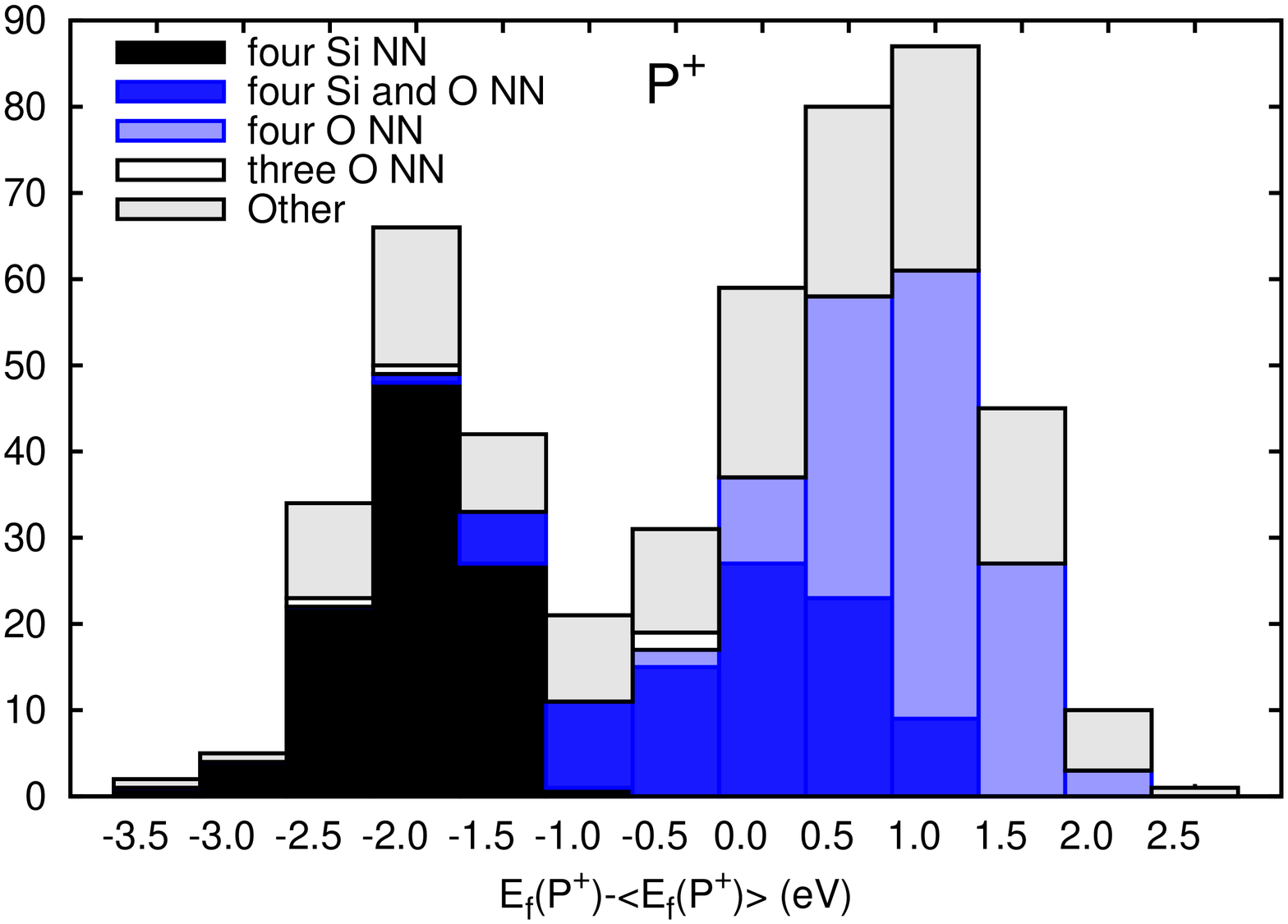}
\caption{
Energy vs. type and number of nearest neighbors (NN) for B$^-$ (top) and P$^+$ (bottom).
The formation energy is given relative to the average of all samples $\langle E_f\rangle$.
The histogram is stacked.
}
\label{fig:types}
\end{figure}

A fundamental factor determining the change of formation energy from the core to the shell
is the difference between the bonding energies of B or P to Si or O.
We analyze the correlation between the formation energy and the type and number
of bonds to the nearest neighbors,
counting the number of bonds in the basis of a geometrical criteria:
if the distance between two atoms exceeds the sum of their
Van der Waals radii by less than 0.5~\AA,
we consider the two atoms bonded\cite{Note}.
The top histogram of Fig.~\ref{fig:types} shows that 
the energy distribution for substitutional boron is asymmetric, with a negative skew.
In general, three-fold coordinated structures have low energy.
Four-fold coordinated structures are evenly distributed on the two sides of the peak,
independently of B$^-$ bonding to Si or O.
This shows that the cost of replacing a Si-Si bond by a B$^-$-Si bond is
approximately the same as the cost of replacing a Si-O bond by a B$^-$-O bond.

For P$_s^+$, the energy distribution is bimodal (Fig.~\ref{fig:types}, bottom).
The lowest energy peak is centered at about 2~eV below the average formation energy
$\langle E_f\rangle$, whilst the higher energy peak is about 1~eV above it.
The lowest energy peak is dominated by four-fold coordinated defects.
Structures where P$_s^+$ bonds to four Si neighbors have the lowest energy,
followed by the structures where P$^+$ is bonded to both O and Si.
The samples where P$^+$ has four oxygen neighbors are highest in energy
amongst four-fold coordinated structures.
Hence, the preference of phosphorus for the Si core is driven
by the lower energy necessary to form a P$^+$-Si bond at the expense
of a Si-Si bond, as compared to the energy necessary to
form a P$^+$-O bond at the expense of a Si-O bond.

We now compare the results of the present investigation
to those found for H-terminated nanocrystals\cite{carvalho-OH}.
When discussing the behavior observed in samples
grown by different methods, we assume that the dopants have enough thermal 
energy to diffuse to the most favorable dopant locations.
During growth, that is very likely the case.

Since H is less electronegative than O, in H-terminated Si nanocrystals
the formation energy of four-fold coordinated B$^-$ or P$^+$ is nearly independent 
on the lattice position it occupies\cite{carvalho-OH}.
Both B and P are stabilized at the surface by loosing one hydrogen to become three-fold coordinated,
but the energy released is greater for the latter.
This may explain why P segregates to the surface of H-terminated silicon nanocrystals,
but the same has not been observed for B\cite{pi-APL-92-123102}.
On the other hand, we find that P should diffuse from the SiO$_2$ shell of SiO$_2$-covered nanoparticles
 or to the Si-SiO$_2$ interface to the Si nanoparticle core.
This is in agreement with the segregation behavior observed for Si nanocrystals embedded in an SiO$_2$
matrix\cite{perego-Nt-21-02560}.
Boron, however, is predicted to have no energetic preference for the Si or SiO$_2$ regions.

\begin{acknowledgments}
The computations were performed on resources provided by the Swedish National Infrastructure for Computing (SNIC),
at KTH (Lindgren) and Ume\aa\ University (Akka),
 University of Aveiro (Blafis) and Milipeia. 
The work was funded by the Calouste Gulbenkian Foundation, the Marie Curie Program PEOPLE (SiNanoTune), FCT Portugal (SFRH/BPD/66258/2009 and PTDC/FIS/112885/2009) and NanoTP.
AC is indebted to Jos\'e Coutinho and Auke Akkerman for useful discussions.
\end{acknowledgments}
%

\begin{thebibliography}{[1]}
\bibitem{takeoka-PRB-62-16820} S. Takeoka, M. Fujii, and S. Hayashi, Phys. Rev. B {\bf 62}, 16820 (2000). 
\bibitem{timmerman-NN-6-710} D. Timmerman, J. Valenta, K. Dohnalov\'a,  W.D.A.M. de Boer, and T. Gregorkiewicz, Nature Nanotechnology {\bf 6}, 710 (2011). 
\bibitem{kovalev-PRL-81-2803} D. Kovalev, H. Heckler, M. Ben-Chorin, G. Polisski, M. Schwartzkopff, and F. Koch, Phys. Rev. Lett. {\bf 81}, 2803 (1998).
\bibitem{pi-APL-92-123102} X. D. Pi, R. Gresback, R. W. Liptak, S. A. Campbell, and U. Kortshagen, Appl. Phys. Lett. {\bf 92}, 123102 (2008).
\bibitem{stegner-PRB-80-165326} A. R. Stegner, R. N. Pereira, R. Lechner, K. Klein, H.~Wiggers, M. Stutzmann, and M. S. Brandt, Phys. Rev. B {\bf 80}, 165326 (2009). 
\bibitem{fuji-in-pavesi-book} Minoru Fuji in {\it Silicon Nanocrystals; Fundamentals, Synthesis, and Applications}, ed. by Lorenzo Pavesi and Rasit Turan (Wiley-VCH, Weinheim, 2010).
\bibitem{cantele-PRB-72-113303} G. Cantele, E. Degoli, E. Luppi, R. Magri, D. Ninno, G. Iadonisi, S. Ossicini, Phys. Rev. B  {\bf 72}, 113303 (2005).
\bibitem{carvalho-OH} A. Carvalho, M. J. Rayson, and P. R. Briddon, J. Phys. Chem. C {\bf 116}, 8243 (2012).
\bibitem{grove-JAP-35-2695} A. S. Grove, O. Leistiko, Jr., and C. T. Sah, J. Appl. Phys. {\bf 35}, 2695 (1964).
\bibitem{ma-APL-98-173103} Jie Ma, Su-Huai Wei, Nathan R. Neale, and Arthur J. Nozik Appl. Phys. Lett. {\bf 98}, 173103 (2011).
\bibitem{xu-PRB-75-235304} Qiang Xu, Jun-Wei Luo, Shu-Shen Li, Jian-Bai Xia, Jingbo Li, and Su-Huai Wei, Phys. Rev. B {\bf 75}, 235304 (2007). 
\bibitem{cole-SS-601-4888} D. J. Cole, M. C. Payne, and L. Colombi Ciacchi, Surf. Sci. {\bf 601}, 4888 (2007).
\bibitem{chang-JAP-103-053517} R. D. Chang and J. R. Tsai, J. Appl. Phys. {\bf 103}, 053517, (2008).
\bibitem{furuhashi-ELEX-1-126} Masayuki Furuhashi, Tetsuya Hirose, Hiroshi Tsuji, Masayuki Tachi, and Kenji Taniguchi, IEICE Electronics Express {\bf 1}, 126 (2004).
\bibitem{jun-ME-89-120} Jun Oh, Hyeon-Kyun Noh, Geun-myung Kim, and Kee Joo Chang, Microelectronic Engineering {\bf 89}, 120 (2012). 
\bibitem{newman} R. C. Newman, in: {\it Early Stages of Oxygen Precipitation in Silicon}, ed. by R. Jones, NATO ASI series (Kluwer Academic Publishers, Dordrecht, 1996), pp. 19.
\bibitem{bracht} H. Bracht, H. H. Silvestri, I. D. Sharp, and E. E. Haller, Phys. Rev. B {\bf 75}, 035211 (2007).
\bibitem{briddon-pssb-217-131} P. Briddon and R. Jones, Phys. Stat. Sol. (b) {\bf 217}, 131 (2000).
\bibitem{rayson-cpc-178-128} M. J. Rayson and P. R. Briddon, Computer Phys. Commun. {\bf 178}, 128 (2008).
\bibitem{rayson-PRB-80-205104} M. J. Rayson and P. R. Briddon, Phys. Rev. B {\bf 80}, 205104 (2009).
\bibitem{briddon-pssb-248-1309}  P. R. Briddon and M. J. Rayson,  Phys. Stat. Sol. (b) {\bf 248}, 1309 (2011).
\bibitem{HGH} C. Hartwigsen, S. Goedecker, and J. Hutter,  Phys. Rev. B 58, 3641 (1998).
\bibitem{goss-TAP-104-69} J. P. Goss, M.J. Shaw, and P.R. Briddon, in: {\it Theory of Defects in Semiconductors}, ed. by D. A. Drabold and S. K. Estreicher, Topics in Applied Physics Vol. {\bf 104} (Springer, Berlin, 2007), pp. 69.
\bibitem{pade} S. Goedecker, M. Teter and J. Hutter, Phys. Rev. B {\bf 54}, 1703 (1996).
\bibitem{nanoSi-EMRS} Alexandra Carvalho, S. \"Oberg, M. Barroso, Mark J. Rayson, and Patrick R. Briddon, to appear in phys. stat.solidi c.
\bibitem{previous} Dong Han, D. West, Xian-Bin Li, Sheng-Yi Xie, Hong-Bo Sun, and S.B. Zhang, Phys. Rev. B {\bf 82}, 155132 (2010).
\bibitem{Note}A more physical criteria should be based on an analysis of the overlap population,
but this would be prohibitive for the number of bonds and samples considered here.
\bibitem{perego-Nt-21-02560} Michele Perego, Caroline Bonafos, and Marco Fanciulli, Nanotechnology {\bf 21}, 25602 (2010).

\end{thebibliography}
%

\end{document}